\documentclass[aps,prapp,twocolumn,showpacs,floatfix]{revtex4}
\usepackage{graphicx,graphics}
\usepackage{natbib}
\usepackage{lipsum}
\usepackage{dcolumn}
\usepackage{amsmath,amssymb,amsfonts}
\usepackage{latexsym,verbatim}
\usepackage{bm}
\usepackage{color}
\usepackage{ulem}
\usepackage{siunitx}
\usepackage[breaklinks=true,colorlinks,citecolor=blue,linkcolor=blue,urlcolor=blue]{hyperref}

\begin{document}
	
\title{First-principles method to study near-field radiative heat transfer}
\author{Tao Zhu}
\email{a0129455@u.nus.edu}
\author{Zu-Quan Zhang}
\author{Zhibin Gao}
\author{Jian-Sheng Wang}
\affiliation{Department of Physics, National University of Singapore, Singapore 117551, Republic of Singapore}

\begin{abstract}
We present a general and convenient first principle method to study near-field radiative heat transfer. We show that the Landauer-like expression of heat flux can be expressed in terms of a frequency and wave-vector dependent macroscopic dielectric function which can be obtained from the linear response density functional theory. A random phase approximation is used to calculate the response function. We computed the heat transfer in three systems -- graphene, molybdenum disulfide (MoS$_2$), and hexagonal boron nitride (h-BN). Our results show that the near-field heat flux exceeds the blackbody limit up to four orders of magnitude.  With the increase of the distances between two parallel sheets, a $1/d^2$ dependence of heat flux is shown, consistent with Coulomb's law. The heat transfer capacity is sensitive to the dielectric properties of materials. Influences from chemical potential and temperature are also discussed. Our method can be applied to a wide range of materials including systems with inhomogeneities which provides solid references for applications of both physics and engineering.
\end{abstract}
\maketitle

\section{Introduction}
Near-field radiative heat transfer (NFRHT) plays an important role in developing novel technologies such as thermal management \cite{thermalmanagement1,thermalmanagement2}, thermal lithography \cite{lithography}, energy conversion \cite{energyconversion1,energyconversion2}, data storage \cite{datastorage1,datastorage2}, and thermophotovoltaic devices \cite{thermophotovoltaic1,thermophotovoltaic2,thermophotovoltaic3}, etc. Both theoretical \cite{theoretical1,volokitin} and experimental \cite{experimental1,experimental2} works have shown that thermal radiation in systems with distances comparable to or smaller than the thermal wavelength $\lambda_T =2\pi\hbar c/(k_BT)$ exceeds the blackbody limit by several orders of magnitude. Theoretically, fluctuational electrodynamics proposed by Rytov \cite{rytov1,rytov2} and further developed by Polder and van Hove \cite{polder} provides a solid and widely recognized description of NFRHT. The energy flux is generated by thermally driven fluctuating electromagnetic fields and the current fluctuations can be characterized by the fluctuation-dissipation theorem (FDT) at the local thermodynamic temperature \cite{rytov1,fdt1,fdt2,fdt3,fdt4}. In this case, the correlation between temperature-driven electrical currents is directly related to the dielectric properties of the materials.

The general framework of fluctuational electrodynamics is macroscopic which combines Maxwell's equations with the FDT of Callen and Welton \cite{fdt2}. The heat flux across a vacuum gap is given by a Landauer-type expression with a transmission function which consists of contributions from both propagating and evanescent waves. The dramatic increase of thermal radiation in the near field is due to the tunneling of evanescent waves which decay exponentially with the gap size. On the other hand, from a microscopic quantum mechanical point of view \cite{mahan}, thermal radiation can be attributed to both Coulomb interactions between charge fluctuations and photonic interactions between transverse current fluctuations. Moreover, Coulomb interactions dominate the energy transfer at the near field and correspond to the evanescent part given by the theory of the fluctuational electrodynamics. 

There are several ways to study the contribution of Coulomb interactions to the energy transfer between two closely separated bodies. One promising approach is using the nonequilibrium Green's function (NEGF) method \cite{negf1,negf2,negf3,negf4,negf5}. A Caroli formula of the transmission function can be obtained from the Meir-Wingreen formula in a local equilibrium approximation. High order many-body effects can also be incorporated in an NEGF method. Another approach is to calculate the net balance of the work done by thermally fluctuating charges in a linear response framework \cite{Yu}. The starting point of this method is to consider the Joule heating effect from charge fluctuations due to external electric fields. In this scheme, the susceptibility function describes the response of internal charge density to a fluctuating external potential and the heat flux can be obtained by averaging thermal fluctuations which are evaluated by FDT. Regardless of the different notations and physical quantities used, equivalence between these two methods has been shown \cite{equivalent}.

In this work, we present a first principle method to investigate the NFRHT problem. We prove that the transmission functions from microscopic quantum mechanical models can be expressed by a formula of a frequency and wave-vector dependent macroscopic dielectric function, consistent with the results of the fluctuational electrodynamics. Moreover, the macroscopic dielectric function can be obtained from the linear response density functional theory (DFT).  The Heat flux of three representative two-dimensional (2D) materials has been studied with a random phase approximation (RPA). Our results show that, at small distances, the heat flux exceeds the traditional Planckian radiative process with several orders of magnitude. Moreover, an asymptotic $1/d^2$ dependence of heat flux is shown with the increase of distances. Our approach is general and can be easily applied to a variety of materials with both homogeneous and inhomogeneous lattice structures.

\section{Method}

\begin{figure}
	\includegraphics[width=8.6 cm]{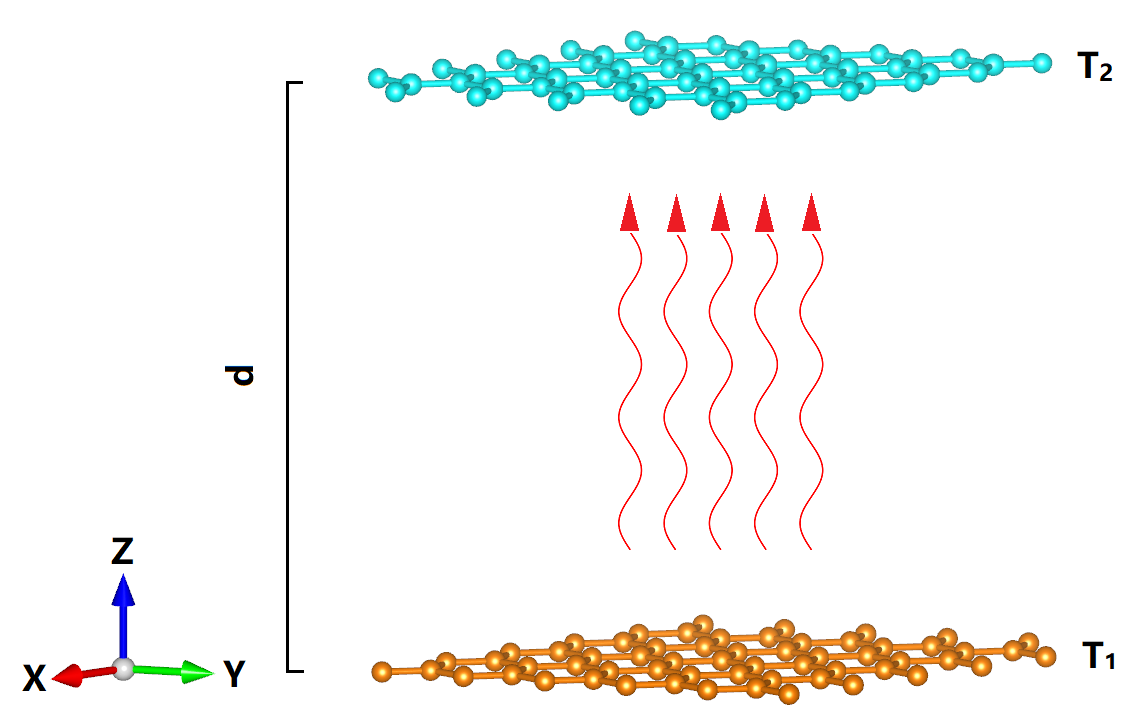}
	\caption{Sketch of radiative heat transfer between two vacuum-gaped 2D materials with a hexagonal lattice. The distance $d$ between the two plates is assumed much smaller than the thermal wavelength $\lambda_T$ at temperature $T_1$ and $T_2$. }
	\label{fig1}
\end{figure}

We consider the heat transfer between two parallel vacuum-gapped 2D plates as sketched in Fig. \ref{fig1}. Plate 1 locates at $z=0$ with temperature $T_1$ and plate 2 is placed at $z=d$ with temperature $T_2$. Each of the plates is in its own internal thermal equilibrium state and net radiative heat transfer between two plates will take place if their temperatures are different. We further assume $T_1 > T_2$ so that heat transfers from plate 1 to plate 2.

\subsection{Transmission function}
Adopting the Coulomb gauge, due to neutrality of induced charge, the Coulomb forces behave as dipole-dipole interactions and can be fully described by a scalar potential. Assuming the interaction is instantaneous (quasi-static limit, the speed of light $c\to\infty$) and neglecting contributions from electromagnetic radiation which is described by a transverse vector potential, the heat flux intensity between two vacuum-gapped bodies is given by a Landauer-like formula \cite{Yu,negf2,negf3,negf4,equivalent}
\begin{equation}
P= \int_{0}^{\infty}\dfrac{d\omega}{2\pi} \hbar\omega \bigl[N_1(\omega)-N_2(\omega)\bigr]S(\omega),
\label{transmission function}
\end{equation}
where $N_{1,2} = 1/(e^{\hbar\omega/(k_BT_{1,2})}-1)$ is the Bose (or Planck) distribution function. The transmission function is given by the Caroli formula \cite{caroli,Yu}
\begin{equation}
S(\omega)=4\,{\rm Tr}\bigl\{\Delta^\dagger_2 \cdot v_{21} \cdot {\rm Im}[\chi_1(\omega)] \cdot v_{12} \cdot \Delta_2 \cdot {\rm Im}  [\chi_2(\omega)]\bigr\},
\label{caroli}
\end{equation}
where 
\begin{equation}
\Delta_2 = \bigl({\rm I}_2 - \chi_2\cdot v_{21}\cdot \chi_1\cdot v_{12}\bigr)^{-1}
\label{delta}
\end{equation}
is the multiple scattering matrix between the two plates. Subscripts 1 and 2 denote quantities of plate 1 and plate 2,  respectively. Exchanging the subscripts does not change the results because of the symmetry. $v_{12,21}$ is the Coulomb interaction between two plates and $\chi_{1,2}$ is the charge density correlation function which describes the response of each plate in terms of induced charge density to the external potential. Tr denotes the trace operation,  ${\rm I}$ is the identity, and the dot indicates convolution or matrix multiplication  depending on how the quantities are represented, e.g., in real space with position ${\bf r}$ or in reciprocal space with lattice vector ${\bf G}$.
   
Considering the translational invariance of quantities in Eq.~(\ref{caroli}) in real space with respect to in-plane lattice vectors, the periodicity of the crystal requiring that the Fourier expansion of a correlation function is defined as
\cite{giulianivignale}
\begin{equation}
f({\bf r},{\bf r}^\prime) = \sum_{{\bf G}, {\bf G}^\prime}\!\int\!\dfrac{d^2{\bf q}}{(2\pi)^2}e^{i({\bf q} + {\bf G})\cdot{\bf r}}f_{{\bf G},{\bf G}^\prime}({\bf q})e^{-i({\bf q} + {\bf G}^\prime)\cdot{\bf r}^\prime},
\label{fourier}
\end{equation} 
here we sum over the set of all reciprocal lattice vectors $\bf G$, ${\bf G}^\prime$, and integrate over $\bf q$ in the first
Brillouin zone. We may check that this expansion satisfies the required lattice translation symmetry, 
$f({\bf r} + {\bf R} ,{\bf r}^\prime + {\bf R}) = f({\bf r},{\bf r}^\prime)$, where $\bf R$ is any real space lattice vector.
We further assume that the two vacuum-gapped sheets have identical lattice constants and the same set of lattice vectors $\bf R$, i.e., we ignore possible lattice mismatch between the two plates so that the expansion is valid for both plates. The convolutions of correlation functions in real space become matrix multiplications in reciprocal space and we can write Eq.~(\ref{caroli}) as
\begin{widetext}
\begin{equation}
S(\omega)=4\int\dfrac{d^2{\bf q}}{(2\pi)^2}\sum_{{\bf G},{\bf G}^\prime, {\bf G}^{\prime\prime},{\bf G}^{\prime\prime\prime}}e^{-\vert{\bf q} + {\bf G}\vert d}{\rm Im}(\varepsilon^{-1}_2)_{{\bf G},{\bf G}^{\prime}}
(\Delta_2^T)_{{\bf G}^{\prime},{\bf G}^{\prime\prime}}e^{-\vert{\bf q} + {\bf G}^{\prime\prime}\vert d}{\rm Im}(\varepsilon^{-1}_1)_{{\bf G}^{\prime\prime},{\bf G}^{\prime\prime\prime}}(\Delta_1^\dagger)_{{\bf G}^{\prime\prime\prime},{\bf G}}
\label{caroli2}
\end{equation} 
with
\begin{equation}
\bigl((\Delta_{2}^{T})^{-1}\bigr)_{{\bf G},{\bf G}^\prime} = \delta_{{\bf G},{\bf G}^\prime} - \sum_{{\bf G}^{\prime\prime}}e^{-\vert{\bf q} + {\bf G}\vert d}(\varepsilon_{1}^{-1}-{\rm I})_{{\bf G},{\bf G}^{\prime\prime}}e^{-\vert{\bf q} + {\bf G}^{\prime\prime}\vert d}(\varepsilon_{2}^{-1}-{\rm I})_{{\bf G}^{\prime\prime},{\bf G}^\prime}.
\label{delta2}
\end{equation}
\end{widetext} 
$(\Delta_{1}^{T})^{-1}$ is obtained from the expression (\ref{delta2}) by swapping $\varepsilon_{1}$ with $\varepsilon_{2}$. The superscript $-1$ means matrix inverse.
 
In arriving at Eq.~(\ref{caroli2}), we have used the relation
\begin{equation}
\left[\varepsilon^{-1}_{\alpha}\right]_{{\bf G},{\bf G}^\prime} = \delta_{{\bf G},{\bf G}^\prime}+\left[v_{\alpha}\right]_{{\bf G},{\bf G}}\left[\chi_{\alpha}\right]_{{\bf G},{\bf G}^\prime}
\label{dyson1}
\end{equation}  
where $\alpha$ denotes 1 or 2, and the Fourier transform of the bare Coulomb interaction in two dimensions is
\begin{equation}
v_{{\bf G},{\bf G^\prime}} = \delta_{{\bf G},{\bf G}^\prime}\dfrac{e^{-\vert {\bf q}+{\bf G}\vert\vert z - z^\prime\vert}}{2\varepsilon_0\vert{\bf q}+{\bf G}\vert}
\label{coulomb}
\end{equation}
where $\varepsilon_0 \approx 8.85\times10^{-12} \,$ F/m is the vacuum permittivity. With this manipulation, we see that we can write the expression of Yu {\sl et al.}\@ \cite{Yu} solely in terms of the inverse of the dielectric function. This is because of the relation, in matrix form, $\varepsilon^{-1}_{\alpha} = I + v_\alpha \chi_\alpha$. 

We adopt the macroscopic approximation, i.e., take $\bf G=G^\prime=G^{\prime\prime}=G^{\prime\prime\prime}=0$. This is an excellent approximation provided that the unit cells are small compared to the distance $d$ and the first nonzero $\bf G$s are large, due to the presence of the exponential factors $e^{-\vert{\bf q} + {\bf G}\vert d}$. Replacing the integral of Eq.~(\ref{caroli2}) by a sum of parallel wave-vectors in the first Brillouin zone, we get
\begin{equation}
S(\omega)=\frac{1}{A} \sum_{{\bf q} \in 1 \textrm{BZ}}\dfrac{4e^{-2qd}\,{\rm Im}(\varepsilon_1^{-1})_{00}\,{\rm Im}(\varepsilon_2^{-1})_{00}}{\left \vert 1-e^{-2qd}\left[(\varepsilon_1^{-1})_{00}-1\right]\left[(\varepsilon_2^{-1})_{00}-1\right]\right \vert^2}
\label{caroli3}
\end{equation}
where $A$ is the area of the sample.

It is worth noting that even though we adopted the macroscopic approximation, contributions from different reciprocal lattice vectors are included in the matrix inversion of $\varepsilon^{-1}_{\bf{G},\bf{G}^\prime}$. The term $(\varepsilon_{1,2}^{-1})_{00}$ as shown in Eq.~(\ref{caroli3}) is the reciprocal of the so called macroscopic dielectric function which will be discussed below. As one may expect, after adopting the macroscopic approximation, Eq.~(\ref{caroli3}) coincides with the evanescent modes of transmission coefficient obtained by fluctuational electrodynamics in the quasi-static limit \cite{polder,volokitin}.

\subsection{Dielectric function}
The microscopic dielectric functions can be obtained from first principles in the framework of the Kohn-Sham density functional theory \cite{ks}. In the random phase approximation, the independent particle polarizability is given by \cite{rpa}
\begin{eqnarray}
\label{chi}
&&\Pi^{0}_{{\bf G},{\bf G}^\prime}({\bf q},\omega)=\dfrac{2e^2}{\Omega}\sum_{n,n',{\bf k}}\textit{w}_{{\bf k}}\bigl(f_{n'{\bf k}+{\bf q}}-f_{n{\bf k}}\bigr)\\
&&\nonumber\times\left(\dfrac{\langle\phi_{n{\bf k}}|e^{-i({\bf q}+{\bf G})\cdot {\bf r}}|\phi_{n'{\bf k}+{\bf q}}\rangle\langle\phi_{n'{\bf k}+{\bf q}}|e^{i({\bf q}+{\bf G}^\prime)\cdot  {\bf r}}|\phi_{n{\bf k}}\rangle}{\epsilon_{n'{\bf k}+{\bf q}}-\epsilon_{n{\bf k}}-\hbar \omega-i\eta}\right),
\end{eqnarray}
where $e$ is electron charge, $\phi_{n{\bf k}}$ and $\epsilon_{n{\bf k}}$ are Kohn-Sham eigenfunctions and eigenvalues, respectively, and ${\bf r}$ is the electron position operator. The damping factor $\eta$ is a small positive quantity that accounts for the broadening of spectra. $\bf{q}$ is the Bloch wave-vector which lies in the first Brillouin zone, $\bf{G}$ and $\bf{G}^\prime$ are reciprocal lattice vectors. The Fermi occupation function $f$ equals 1 for occupied states and 0 for unoccupied states. $\Omega$ is the area of the primitive cell. $\textit{w}_{{\bf k}}$ is the weight of each $k$-point in the first Brillouin zone which is defined to sum to one and the factor 2 accounts for the spin degeneracy.  Equation (\ref{chi}) shows that the Kohn-Sham response function is a summation of independent transitions from the filled to the empty states \cite{Onida,lrpdft}.

Then the Kohn-Sham microscopic independent particle dielectric matrix is given by
\begin{equation}
\label{epsilon}
\varepsilon_{{\bf G},{\bf G}^\prime}({\bf q},\omega)=\delta_{{\bf G},{\bf G}^\prime}-v_{{\bf G},{\bf G}}({\bf q})\Pi^{0}_{{\bf G},{\bf G}^\prime}({\bf q},\omega),
\end{equation}
where $v$ is the Fourier transform of the bare Coulomb interaction.

The macroscopic dielectric function is determined as \cite{adler,wiser}
\begin{equation}
\varepsilon_M(\bf q, \omega)=\dfrac{1}{(\varepsilon^{-1})_{{\bf G}=0,{\bf G}^\prime=0}(\bf{q},\omega)}.
\label{macro}
\end{equation}
The off-diagonal elements of the microscopic dielectric function in the matrix inversion of Eq.~(\ref{macro}), $\varepsilon^{-1}_{\bf{G},\bf{G}^\prime}(\bf{q},\omega)$, are responsible for the so-called local field effects and they become important in systems with inhomogeneous lattice structures \cite{lfe,lfe1,lfe2,lfe3,lfe4,lfe5,lfe6}. However, for most materials with near-homogeneous charge distribution, we can write Eq.~(\ref{macro}) in an independent particle form:
\begin{equation}
\varepsilon_M(\bf q, \omega)=\varepsilon_{0,0}(\bf{q},\omega)
\label{nlf}
\end{equation}
where the off-diagonal elements of the matrix inversion of Eq.~(\ref{macro}) are neglected. For simplicity, we adopted this approximation in our calculation of dielectric functions of homogeneous 2D materials.

\subsection{Computational Details}
As representative cases, we study the near-field heat transfer of vacuum-gaped parallel films of 2D materials. Three typical 2D materials, namely, graphene, molybdenum disulfide (MoS$_2$), and hexagonal boron nitride (h-BN) have been selected to be studied. They are well-known 2D materials with similar hexagonal structures, but their electronic properties are significantly different, i.e., graphene is a semimetal, MoS$_2$ is a semiconductor, and h-BN is an wide-gap insulator. We first calculated their ground state properties by using DFT as implemented in the {\sc Quantum ESPRESSO} \cite{qe1,qe2}. The projector-augmented wave (PAW) method \cite{paw} and the Perdew-Burke-Ernzerhof (PBE) exchange-correlation functional with generalized gradient approximation (GGA) \cite{gga} were employed.  A plane-wave basis set with 40 Ry energy cut-off was used to expand the Kohn-Sham wave functions. The first Brillouin zone was sampled by a $90\times90\times1$ Monkhorst-Pack \cite{mp} grid. The Fermi-Dirac smearing was adopted to treat the partial occupancies for graphene with different temperatures. For example, 1000 K corresponds to a smearing width of 0.0063 Ry. The in-plane lattice constants are $a = b = \SI{2.46}{\angstrom}$ for graphene, $a = b = \SI{3.16}{\angstrom}$ for MoS2, and $a = b = \SI{2.48}{\angstrom}$ for h-BN. To avoid interactions from neighboring lattice in $z$ direction, a large lattice constant of $c = \SI{18}{\angstrom}$ was set to the $z$ direction of the unit cell.

Then the frequency and wave-vector dependent dielectric function is calculated on top of the ground state calculations using the package {\tt Yambo} \cite{yambo1,yambo2}. The frequency cut-off was set to \SI{1}{\electronvolt} which is sufficient for radiative heat transfer calculation. The damping factor $\eta$ is \SI{3}{\milli\electronvolt}, \SI{70}{\milli\electronvolt}, and \SI{250}{\milli\electronvolt} for graphene, MoS$_2$ and h-BN, respectively. These values are corresponding to their electron relaxation lifetimes which are further determined by their electron mobilities, i.e., $\mu \sim$ 5000 cm$^2$/(V$\cdot$s) for graphene \cite{graphene}, $\sim$ 200 cm$^2$/(V$\cdot$s) for MoS$_2$ \cite{mos21}, and $\sim$ 50 cm$^2$/(V$\cdot$s) for h-BN \cite{hbn}.

\section{Results}\label{Results}

\begin{figure}
	\includegraphics[width=8.6 cm]{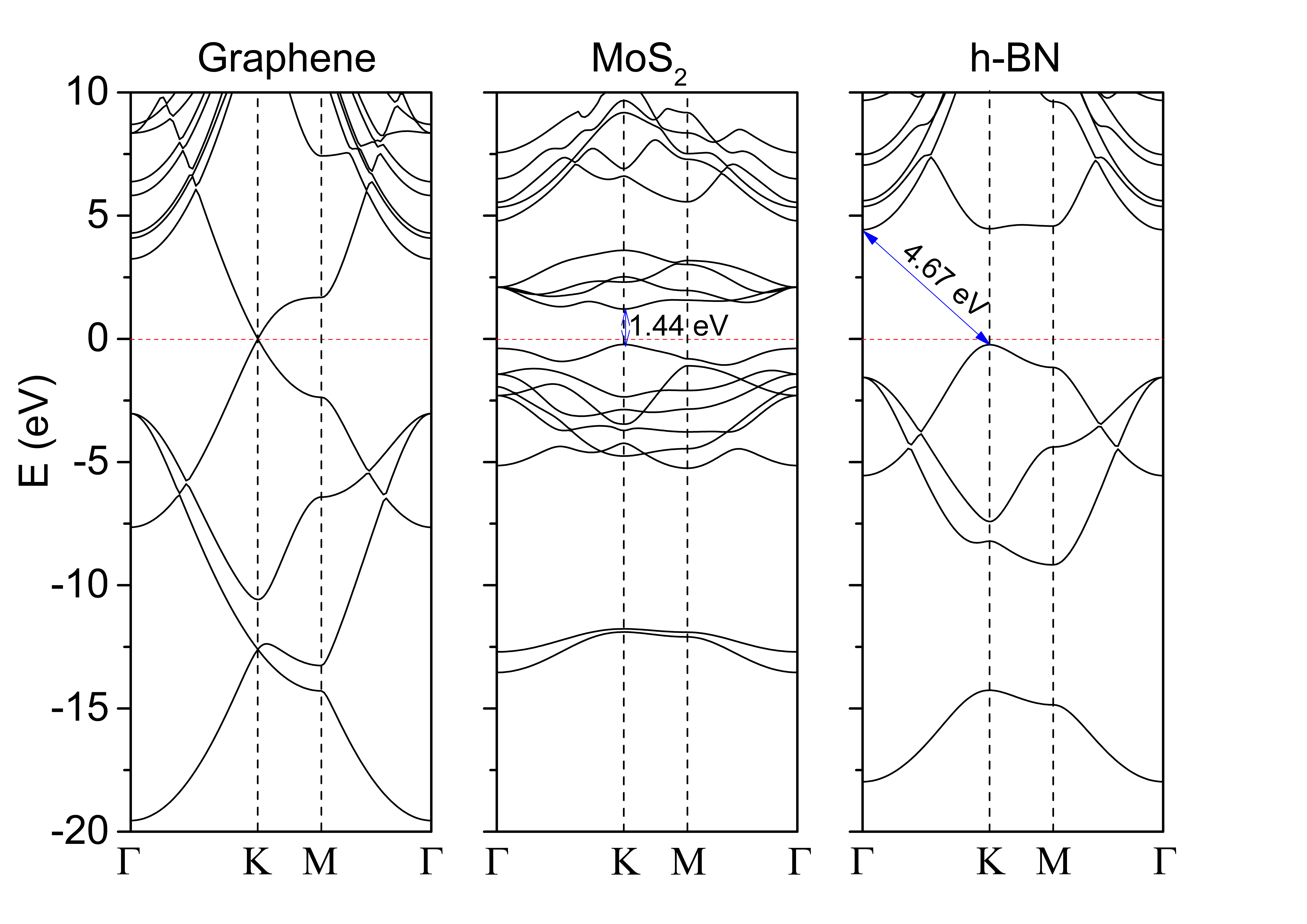}
	\caption{Calculated band structure of single-layer graphene, MoS$_2$, and h-BN along high-symmetry points of the Brillouin zone. Red dashed line represents the Fermi level.}
	\label{fig2}
\end{figure}
We first look at the electronic properties of three studied 2D materials. The calculated band structures of graphene, MoS$_2$, and h-BN along the lines connecting high-symmetry points of the Brillouin zone are shown in Fig.~\ref{fig2}. As can be seen, single-layer graphene displays a semimetallic character as its conduction and valence bands meet at the Dirac point located at the $K$ high-symmetry point of the Brillouin zone. Another Dirac point placed at $K^\prime$ which accounts for the valley degeneracy is not shown here. The cases of MoS$_2$ and h-BN are different. A direct bandgap of 1.44 eV at the $K$ point is present for monolayer MoS$_2$ which indicates that this material is a semiconductor. This feature is different from reported character of its bulk form which the conduction band minimum moves to the $\Gamma$ point and an indirect bandgap is shown \cite{mos22}. On the other hand, single-layer h-BN is an insulator which has a wide indirect bandgap of 4.67 eV with the conduction band minimum lies at the $\Gamma$ point and the valance band maximum is at the $K$ point.

It may be worth noting that the calculations are based on a pure density functional theory with GGA. The calculated bandgap is underestimated to some degree compared to experimental results \cite{mos22,hbn2} and further advanced techniques such as hybrid functional \cite{hybrid} or GW \cite{gw} method will generally correct this problem to a certain extent. However, these methods are rather time-consuming and, more importantly, general features discussed here are correct within the pure DFT.

\begin{figure}
	\includegraphics[width=8.6 cm]{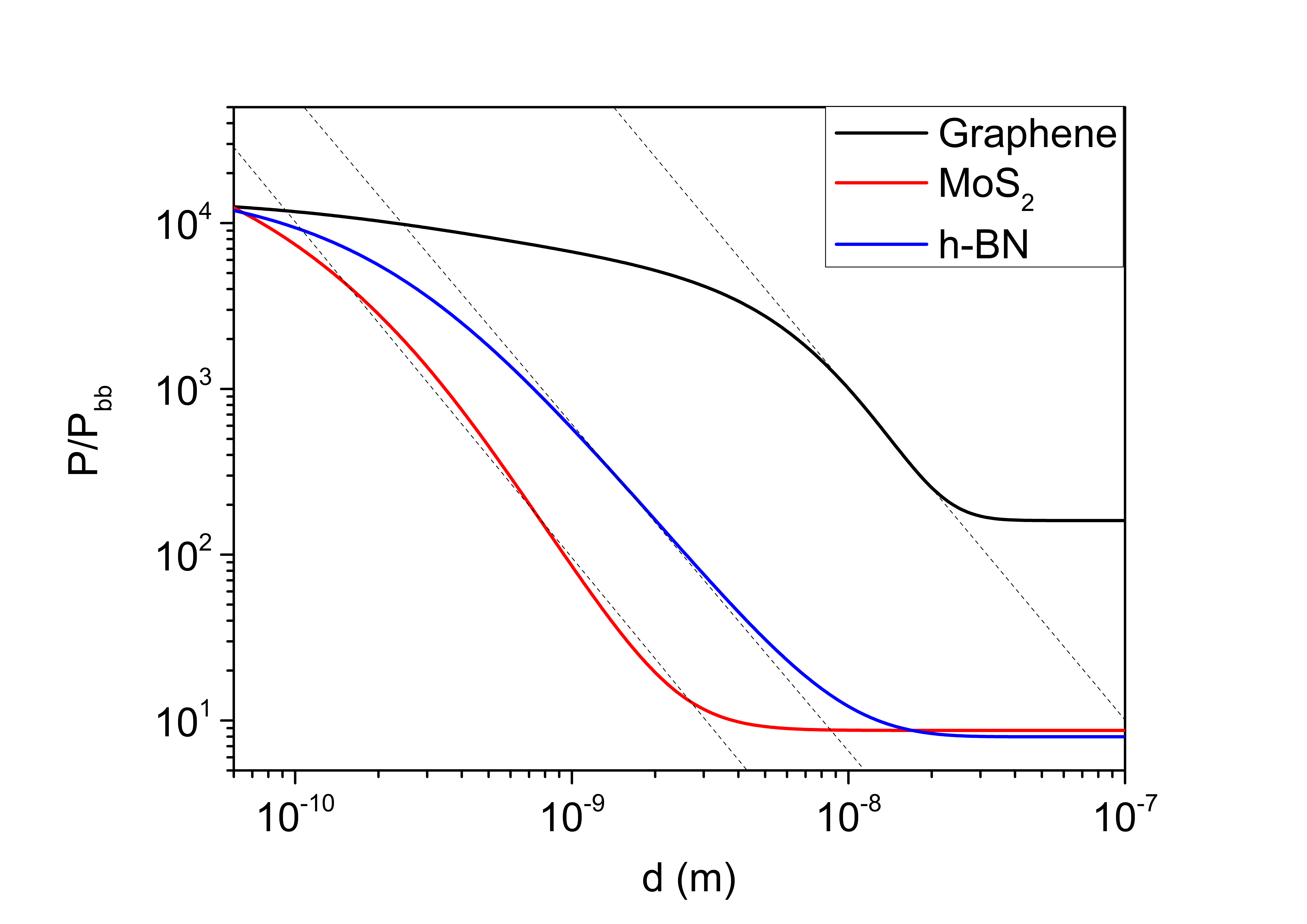}
	\caption{Near-field heat flux ratio of graphene, MoS$_2$, and h-BN with $T_1$ = 1000 K and $T_2$ = 300 K. $P_{bb}$ is the black body heat flux calculated by the Stean-Boltzmann law. The dashed lines have a slope of -2 to show the distance dependence of heat flux.}
	\label{fig3}
\end{figure}

Figure \ref{fig3} shows the calculated heat flux ratio of three studied materials. The horizontal coordinate represents the distance $d$ between two plates and the vertical coordinate is the ratio of calculated near-field heat flux in our model to the heat flux described by the Stefan-Boltzmann law $P_{bb} = \sigma (T_1^4 - T_2^4)$, where $\sigma \approx 5.67\times10^{-8} \,$W/(m$^2$K$^4$) is the Stefan-Boltzmann constant. The expression of $P_{bb}$ is obtained by integrating spectral density in the Planck's law of radiation over the frequency and then calculating the net power radiated between two plates. As we can see, at small distances, the calculated near-field heat flux is 1 $\sim$ 4 orders higher than the results from the blackbody radiation for all three materials.

There are many interesting features shown in Fig.~\ref{fig3}. Generally, the heat flux ratio is decreasing with increasing distances. On the one hand, a convergent value of heat flux ratio $\sim10^4$ is shown for both materials at the leftmost of Fig.~\ref{fig3}. In this case, the distance $d$ is smaller than \SI{1}{\angstrom} which reaches the contact limit so that two plates actually ``touch" each other and no longer satisfy the near-field requirement. One can expect that the results are reliable only if the distance between two sheets is larger than the lattice constant of materials. However, both materials show a convergent trend at small distances up to several angstroms. This agrees with the saturation of the p-polarized evanescent heat flux at small distances when nonlocal effects are taken into account \cite{Chapuis}. The most distinct saturation of heat flux is shown for graphene which shows a relatively high heat flux ratio up to 3 nm. This feature may arise from the rich plasmonic properties of graphene \cite{Chapuis,grapheneplasmon}. Beyond these distances, the heat flux ratio has a distance dependence around $1/d^2$ which is consistent with Coulomb's law and agree to the character of the p-polarized heat flux \cite{negf3,Joulain,volokitin,pablo}. For example, this character is shown for MoS$_2$, h-BN, and graphene at the range around 0.3 $\sim$ 3 nm, 0.6 $\sim$ 18 nm, and 6 $\sim$ 30 nm, respectively.  

On the other hand, for all three materials, the heat flux ratio tends to a constant for distances exceed certain values. For example, the heat flux ratio becomes constant for distances beyond $\sim$3 nm, $\sim$18 nm, and $\sim$30 nm for MoS$_2$, h-BN, and graphene, respectively. In general, the heat flux will become constant at far field ($> 1\, \mu$m) \cite{volokitin} where contributions from propagating waves dominate. However, our theory includes only the longitudinal component of the electromagnetic field (i.e. the Coulomb field) and ignored completely the retardation effect of the transverse field. So, we can not recover a constant at large distances in theory. The ``premature" constancy of heat flux as shown in Fig. \ref{fig3} is not physical and is due to our computational limitations. Firstly, it occurs because the value of transmission coefficient $S(\omega)$ in Eq.~(\ref{caroli3}) decays exponentially with the increase of distance $d$ for all non-vanishing wave-vectors. Thus, for larger distances, the only contribution is the transition at long-wavelength limit ${\bf q} \to 0$ which causes the transmission coefficient a constant. As the nonlocal modes of large $\bf q$ become less important at large distances \cite{Chapuis}, the validity of the Eq.~(\ref{caroli3}) with quasi-static approximation are expected at distances of 100 nm to 1$\, \mu$m. This constancy problem can be partially resolved by increasing the density of k-point sampling in the first Brillouin zone so that smaller non-vanishing $\bf q$ contributes to the sum and further extend the trend to larger distances. This is further supported by preceding work that used the rotational symmetry and transforming the 2D k-point sampling into a one-dimensional problem with much denser k-points \cite{negf3}. Secondly, we study the energy transfer mediated by the Coulomb interaction with the quasi-static approximation. This corresponds to the p-modes of the evanescent waves in the traditional theory of fluctuational electrodynamics. So, at far field, the evanescent waves vanished and the thermal radiation can only be achieved via propagating waves \cite{fdt3}. In this case, our theory is not valid and heat flux at the far field is a constant which is given by the Stefan–Boltzmann law for black bodies \cite{pablo}.

From Eq.~(\ref{caroli3}), the near-field heat transfer between two closely spaced plates relates to the dielectric properties of each material. Meanwhile, the conductivity which is determined by both charge density and mobility also plays a significant role in the NFRHT. This is because the transferred energy is linked to the response function $\chi$ which describes the induced charge density to fluctuating Coulomb interactions. We can expect that graphene has the highest heat transmission among all three investigated materials as it is metallic and has extremely large carrier mobility \cite{graphene}. It is indeed the case as shown in Fig.~\ref{fig3} that graphene has the highest heat transmission among all three materials at all distances. Meanwhile, h-BN shows a relatively higher heat flux ratio than that of MoS$_2$ because the former has larger dielectric functions for most wave-vectors \cite{hbn,hbn2}.

\begin{figure}
	\includegraphics[width=8.6 cm]{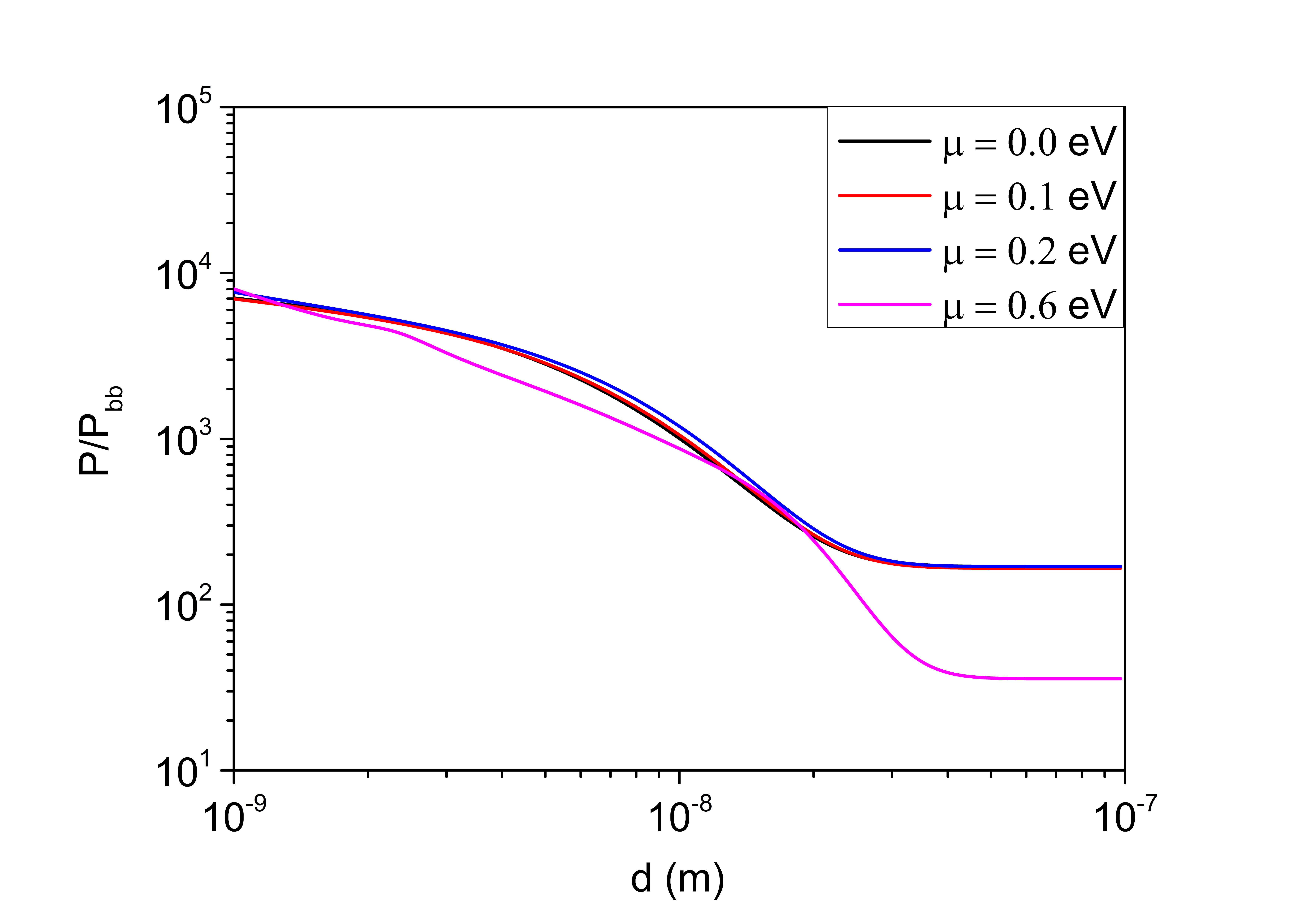}
	\caption{Calculated heat flux ratio of graphene with different chemical potentials. Both plates have the same doping level with $T_1$ = 1000 K and $T_2$ = 300 K.}
	\label{fig4}
\end{figure}

We have shown that graphene is a semimetal whose Fermi level crosses the Dirac point. However, experimental works often measure properties of graphene with substrates or doping which introduces extra chemical potential shifts to the material. Now we study the effect of doping on the near-field heat flux ratios of graphene. The calculated results are shown in Fig.~\ref{fig4}. The chemical potential $\mu$ is the Fermi energy difference between calculations with and without doping. The doping is achieved by introducing extra charges to the system. For example, $\mu =$ 0.1 eV, 0.2 eV, and 0.6 eV correspond to additional 0.0005, 0.001, and 0.005 unit of electrons introduced to one unit cell of graphene. The only difference between two separated graphene sheets is their temperatures, i.e., 1000 K and 300 K, respectively. As can be seen, the heat flux ratio in all doping levels converge to values of $\sim$$10^4$ at small distances and almost identical results are shown for the heat flux ratio with small doping levels. However, for a relatively large doping extent ($e.g.$, 0.6 eV), the heat flux ratio has a lower-lying arch at the range of 1 nm to 10 nm. This is because the doping opens the interband transition gaps. After 10 nm, the heat flux ratios again exhibit a $1/d^2$ character with respect to distance. The relatively lower arch of heat flux ratio at small distances forms the so-called ``doping bubble" as reported in preceding work \cite{negf3}.     

\begin{figure}
	\includegraphics[width=8.6 cm]{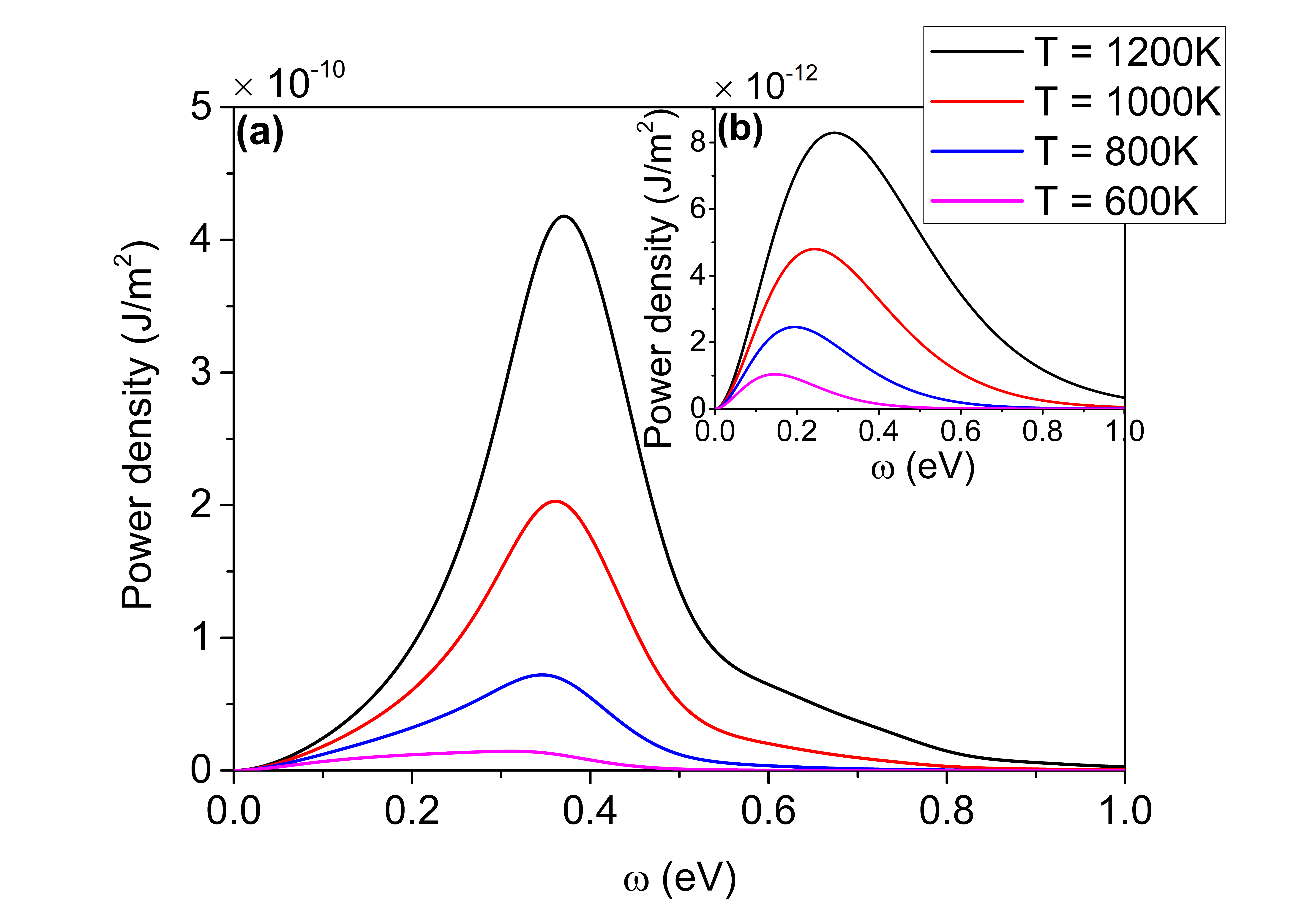}
	\caption{Temperature dependence of power density spectrum of (a) near-field radiation of graphene at 10 nm, and (b) results from Planck's law of blackbody radiation.}
	\label{fig5}
\end{figure}

At last, we discuss the effects from temperature. Fig.~\ref{fig5} is the calculated near-field radiative power density spectrum of graphene at $d = 10$ nm. With the increase of temperature, the power density gradually increases and the characteristic frequency at which spectrum is peaked also shifts to higher frequencies. This is consistent with the spectrum of blackbody radiation which is given by Planck's law as shown in the inset graph of Fig.~\ref{fig5}. However, the near-field radiative power density is approximately 100 times higher than that of the blackbody radiation. Moreover, the characteristic frequency of near-field radiation is also blue-shifted compared to the corresponding results from Planck's law. Besides, we can see that, for the selected temperature range, the spectrum gradually vanishes at high frequency ($\geq$ 1 eV) and thus we set 1 eV as the frequency cut-off in our calculation of the heat transmission function.
    
\begin{figure}
	\includegraphics[width=8.6 cm]{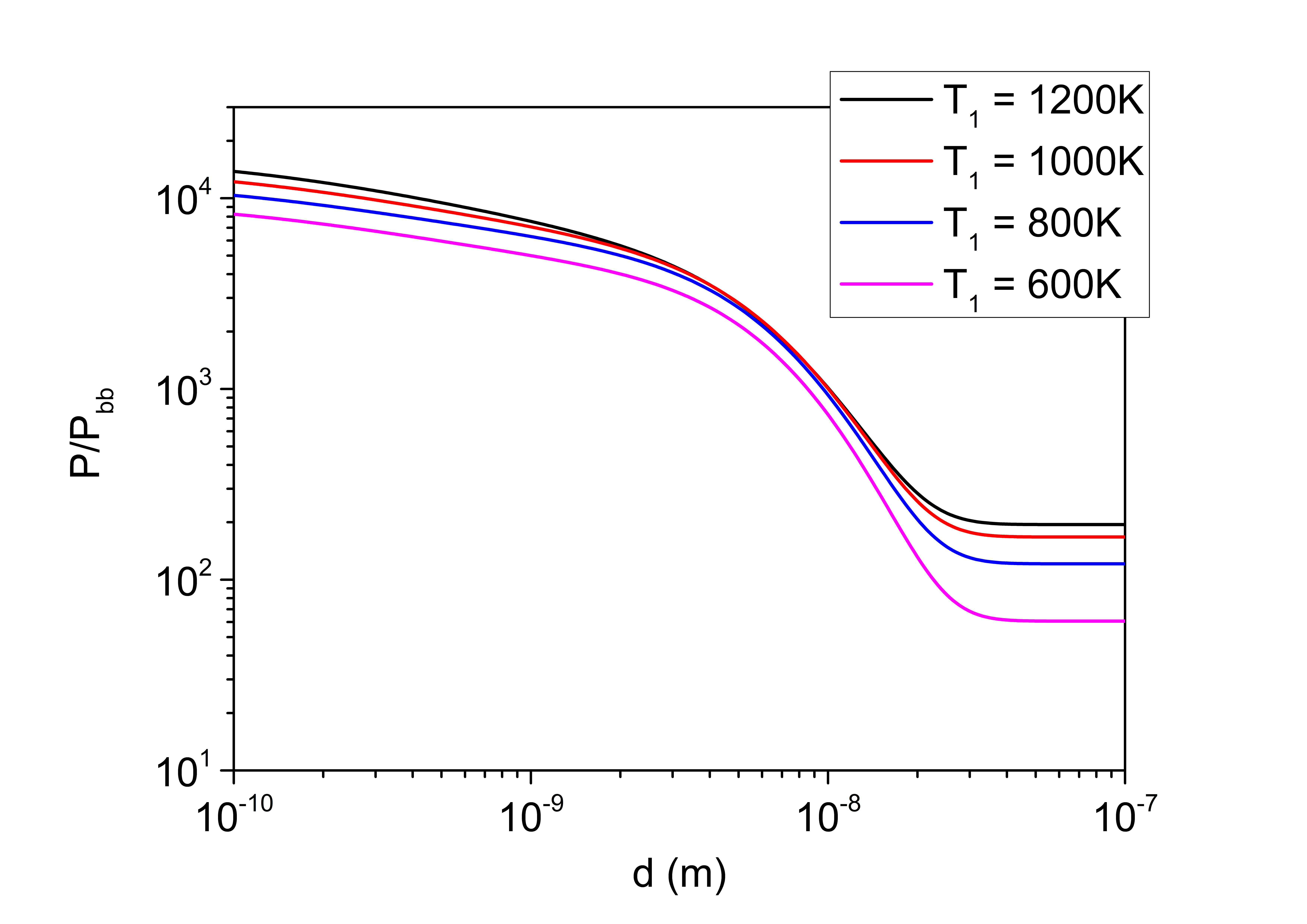}
	\caption{Calculated heat flux ratio of graphene with different temperatures. The temperature of plate 2 is fixed at 300 K and the temperature of plate 1 varies from 600 K to 1200 K.}
	\label{fig6}
\end{figure}

In Fig.~\ref{fig6}, we show the heat flux ratio of two graphene sheets with different temperatures. Without loss of generality, we fix the temperature of sheet 2 at 300 K and change the temperature of sheet 1. As shown, general features are similar to the aforementioned discussions but the heat flux ratio increases with increasing temperatures. This is because the temperature dependence of the Bose function of Eq.~(\ref{transmission function}) is exponential while the denominator $P_{bb}$ depends on the fourth power of the temperature. However, this effect is not significant. A convergence of heat flux ratio is shown at $d \approx 6$ nm for high temperatures which indicates that the dramatic increase of thermal radiation in the near-field is not very sensitive to the temperature.

\section{Discussions}\label{Discussion}

We have shown that our first principles method can be used to study the NFRHT and the results are consistent with preceding reports. Discussions on the possible extensions, limitations, as well as applications of our methods are meaningful.

The present model and expressions introduced above are only applied to monolayer 2D materials with identical lattices on both sides. We further assume that both sides are at its internal thermal equilibrium and we study the energy transfers stemming from the Coulomb interactions. However, further extensions of the current model are promising. For example, radiative heat transfer between multiple plates with finite thickness and out of thermal equilibrium has been studied \cite{latella}. The energy transfers are given by a similar Landauer-like formulation and the transmission coefficients can be expressed in terms of reflection and transmission properties of the different layers. Similarly, in the first principle method, one can utilize the response function in terms of the polarizability $\Pi$ or susceptibility $\chi$ to build the transmission function. Then the calculated response function would be a block diagonal matrix which each non-zero blocks corresponding to the polarizability or the susceptibility of each layer and the off-diagonal interaction terms would be set to zero. This method also gets rid of the requirement that the lattice constant of two sides must match exactly. However, the most notable limitation of these extensions is that a large supercell may be required to calculate the response function matrix that requires substantial computational resources.

On the other hand, we have adopted the quasi-static limit when considering the Coulomb interactions originate from the scalar potential $\varphi$. In this regard, our model only contains the $p$ polarization modes which are most important for the near-field heat transfer of 2D metals like graphene. Meanwhile, to study possible different behaviors of metals and polar materials at very small distances, contributions of $s$ polarization from the vector potential \textbf{A} should be considered. In this case, one can consider the contributions from both propagating waves and evanescent waves with different wave-vectors. Moreover, phonons play an important role in polar materials due to the explicit charges of atoms which strongly interact with the scalar field. These would be interesting and promising extensions of our model in future implementation.

As we mentioned, the heat transfer between two closely separated bodies is a fundamental problem but has been proved to play a pivotal role in applications of many novel technologies in both physics and engineering. Our current method provides a very practical way to compute the near-field heat flux because the transmission function of Eq. (\ref{caroli3}) is built by means of the macroscopic dielectric functions which can be directly obtained from many first principle packages. Conventionally, to calculate the transmission coefficient, one need to do a case-by-case theoretical modeling to get closely related quantities like reflection coefficients $r$ \cite{polder,volokitin}, conductivity $\sigma$ \cite{pablo,ilic,falkovsky1,falkovsky2}, dielectric function $\varepsilon$ \cite{pedersen}, polarizability $\Pi$ \cite{negf3,negf4,guinea,sarma,svintsov}, or susceptibility $\chi$ \cite{Yu}, etc. For example, the free-electron Drude model for metal and the Dirac model for graphene has been widely used to study NFRHT. Nevertheless, simple analytical expressions of these quantities may not be practical for materials with complex structures and electronic configurations. On the contrary, all these quantities can be calculated using the first principle method, and thus one can expect that our method can be applied to many kinds of materials. This is important for practical applications in engineering. Moreover, we believe that our approach is accurate because we directly used the electron density and the Kohn-Sham wavefunctions to built the polarizability and dielectric functions. We can also use approximations beyond RPA and taken into account the local field effect which is very important for inhomogeneous materials. Thus, we believe that our method provides a practical, convenient, and accurate theoretical prediction of near field heat flux and can be widely used in applications.

\section{Summary}
We have studied the near-field radiative heat transfer of vacuum-gapped 2D crystal lattices using a first principle method. The heat flux between two closely separated plates is given by a Landauer-like formula. We have shown that the transmission function can be expressed in a form of macroscopic dielectric functions with summation over all parallel wave-vectors in the first Brillouin zone.  The random phase approximation has been used to calculate the frequency and wave-vector dependent dielectric functions in a linear response density functional scheme.

As representative cases, we investigated the electronic properties and thermal radiation of three typical 2D materials. Our calculations show that the near-field heat fluxes exceed the blackbody limit with up to 4 orders of magnitude. Graphene has the largest heat flux ratio among all three materials because of its higher electron density and mobility. The heat flux ratio exhibits a $1/d^2$ character when the distance between two plates exceeds some extent. A ``doping bubble" is shown in graphene with large chemical potentials. Moreover, the near-field radiation spectrum has similar characters as blackbody radiation. Both power density, character frequency, as well as heat flux ratio have a positive correlation with the temperature. However, the near-field power density is significantly higher than that of Planck's law. All consistent with preceding reports. Improvements can be made to go beyond RPA and single-layer system. With the summations of reciprocal lattice $\bf G$ going beyond just the origin, the method then can handle highly inhomogeneous systems, such as surfaces terminated with an edge or systems with finite thickness using super-cells. Finally, our method is general and can be applied to study near-field radiative heat transfer of various kinds of materials which provides a solid reference for applications of both physics and engineering.

\section{acknowledgments}

This work is supported by a FRC grant R-144-000-402-114 and an MOE tier 2 grant R-144-000-411-112. T.Z. and Z.-Q.Z. contribute equally to this work.

\end{document}